\makeatletter \def\namedlabel#1#2{\begingroup
  \def\@currentlabelname{#2}%
  \label{#1}\endgroup
} \makeatother

\newcommand{\papertitle}{Observation of quantum spin noise in a 1D
  light-atoms quantum interface}
\newcommand{\paperkeywords}{Quantum Optics, Nanofiber, QND
  measurement}

\documentclass[twocolumn,amsmath,amssymb, aps, showpacs, showkeys,
nobibnotes, floatfix, pra, 10pt]{revtex4}

\usepackage[utf8]{inputenc} % Allow for UTF8-characters in TeX-code and bibtex-file
\usepackage[english]{babel} % english language
\usepackage{csquotes} % recommended by babel-package: quotation marks
\usepackage{calc}

\usepackage{graphicx}% Include figure files
\usepackage{siunitx}%
\DeclareSIUnit\gauss{G} % Gauss is not SI
\usepackage{bm}% bold math
\usepackage{xspace}%
\usepackage{xpatch}% http://ctan.org/pkg/xpatch

\newcommand{\linkcolor}{blue}%
\usepackage{hyperref} %% optional
\hypersetup{colorlinks=true, %
  linkcolor=\linkcolor, %
  citecolor=\linkcolor, %
  filecolor=\linkcolor, %
  urlcolor=\linkcolor, %
  pdftitle=\papertitle, %
  pdfauthor={J.-B. Béguin, J. H. Müller, J. Appel, E. S. Polzik}, %
  pdfsubject={Quantum Optics, Quantum noise, Quantum state engineering
    and measurements, Nanofiber}, %
  pdfkeywords=\paperkeywords %
} %

\graphicspath{{images/}}

\makeatletter%
\def \@labelsection{%
  \empty%
}%
\def \@labelsubsection{\@labelsection.\thesubsection}%
\def \@labelsubsubsection{\@labelsubsection.\thesubsubsection}%
\xpatchcmd{\@sect@ltx}{\@xsect}{%
  \let\@hskip\hskip%
  \def \hskip { \@hskip 0em plus}%
  \let\@MakeTextUppercase\MakeTextUppercase%
  \def \MakeTextUppercase{}%
  \edef \@currentlabelname{%
    \@hangfrom@section{}{\csname @label#1\endcsname}{#8}%
  } %
  \let\MakeTextUppercase\@MakeTextUppercase%
  \let\hskip\@hskip%
  \@xsect}{}{}% Patch \section
\makeatother

\newcommand{\var}[1]{\operatorname{var}\left(#1\right)}% variance

\newcommand{\ket}[1]{\ensuremath{|#1\rangle}\xspace}%

\newcommand{\Natom}{N_{\text{at}}}%

\newcommand{\fref}[2][]{Fig.~\ref{#2}\textcolor{\linkcolor}{#1}} % Reference to a figure
\newcommand{\sref}[1]{\nameref{#1}}%
\newcommand{\mean}[1]{\left \langle #1 \right \rangle}% mean

\newcommand{\NBI}{QUANTOP, Niels Bohr Institute, University of
  Copenhagen, Blegdamsvej 17, 2100 Copenhagen, Denmark}

\begin{document}

% \preprint{APS/123-QED}

\newcommand{\correspondingauthors} { %
  \email[Corresponding Authors:~]{polzik@nbi.dk}%
  \email[]{jappel@nbi.dk} %
}

\title{\papertitle}

\author{J.-B. Béguin}%
\altaffiliation[Present address: ]{Norman Bridge Laboratory of Physics 12-33,
  California Institute of Technology, Pasadena, CA 91125, USA}%

\author{J. H. Müller}%
\author{J. Appel} \correspondingauthors%
\author{E. S. Polzik} \correspondingauthors%
\affiliation{\NBI}
\date{\today}

\begin{abstract}
  We observe collective quantum spin states of an ensemble of atoms in
  a one-dimensional light-atom interface. Strings of hundreds of
  cesium atoms trapped in the evanescent fiel of a tapered nanofiber
  are prepared in a coherent spin state, a superposition of the two
  clock states. A weak quantum nondemolition measurement of one
  projection of the collective spin is performed using a detuned probe
  dispersively coupled to the collective atomic observable, followed
  by a strong destructive measurement of the same spin projection. For
  the coherent spin state we achieve the value of the quantum
  projection noise $\SI{40}{\deci\bel}$ above the detection noise,
  well above the $\SI{3}{\deci\bel}$ required for reconstruction of
  the negative Wigner function of nonclassical states. We analyze the
  effects of strong spatial inhomogeneity inherent to atoms trapped
  and probed by the evanescent waves. We furthermore study temporal
  dynamics of quantum fluctuations relevant for measurement-induced
  spin squeezing and assess the impact of thermal atomic motion. This
  work paves the road towards observation of spin squeezed and
  entangled states and many-body interactions in 1D spin ensembles

  \ifdefined\svnid {%
    \begin{description}%
    \item[SVN] \footnotesize%
      \textcolor{red}{\svnFullRevision*{\svnrev} by
        \svnFullAuthor*{\svnauthor}, %\newline %
        Last changed date: \svndate }%
    \end{description}%
  } \fi%
\end{abstract}

% PACS, the Physics and Astronomy
% Classification Scheme.  42.50.Ct Quantum description of interaction
% of light and matter; related experiments 37.10.Jk Atoms in optical
% lattices 42.50.Ex Optical implementations of quantum information
% processing and transfer

% Use showkeys class option if keyword
% display desired
\maketitle

Quantum superpositions, squeezed states and multipartite entanglement
are features of the quantum world which are central for the future
developments of quantum information science and metrology.  These
fundamental nonclassical phenomena have been extensively studied and
exploited with an efficient quantum interface between light and
ensembles of atoms~\cite{revEugene}, offering an exciting alternative
to cavity quantum electrodynamics with single particles.

One of the most recent and promising examples of such light-atom
interface uses an optical nanofiber. Its essence is in combining
guiding of light and trapping of cold atoms in the sub-wavelength
evanescent field of a tapered fiber. This gives rise to strong
light-atom interactions in a genuine one-dimensional geometry
\cite{Vetsch,Kimble, JB, AokiNanofiber}.

Among the recent successes of this emerging platform are generation of
highly sub-Poissonian atom number distributions \cite{JB}, realization
of EIT-based memories \cite{ArnoEIT,LauratEIT}, achieving the strong
cavity-QED regime for a single atom \cite{AokiNanofiber}, and
demonstration of nano-light mirrors with a few structured atoms
\cite{UsMirror,LauratMirror}. These new functionalities become
enabling tools to implement exciting proposals for studies of
one-dimensional many-body effects \cite{njpChang}.  The intrinsic
fiber compatibility of this interface makes it particularly attractive
for scalable quantum fiber networks.

In order to open the realm of continuous variable information
processing and quantum metrology with nanofiber interfaces two well
recognized objectives have to be pursued. The first one is the ability
to characterize the preparation of collective quantum states of atoms
with high-precision tomography \cite{Schmied2011,Sewell2013}. The
second challenging goal is the realization of a quantum non-demolition
(QND) measurement on collective atomic states which is one way for
creation of squeezed states \cite{pnasAppel,Qi2016}, with applications to
entanglement-assisted atomic clocks and magnetometers as well as a
component in quantum teleportation and memory protocols
~\cite{revEugene}.

In this work, we demonstrate for the first time quantum state
tomography and a quantum non-demolition measurement in a nanofiber
trapped atomic ensemble at the projection noise level.

\section{Atomic population measurement}

\begin{figure}
  \includegraphics[width=\columnwidth]{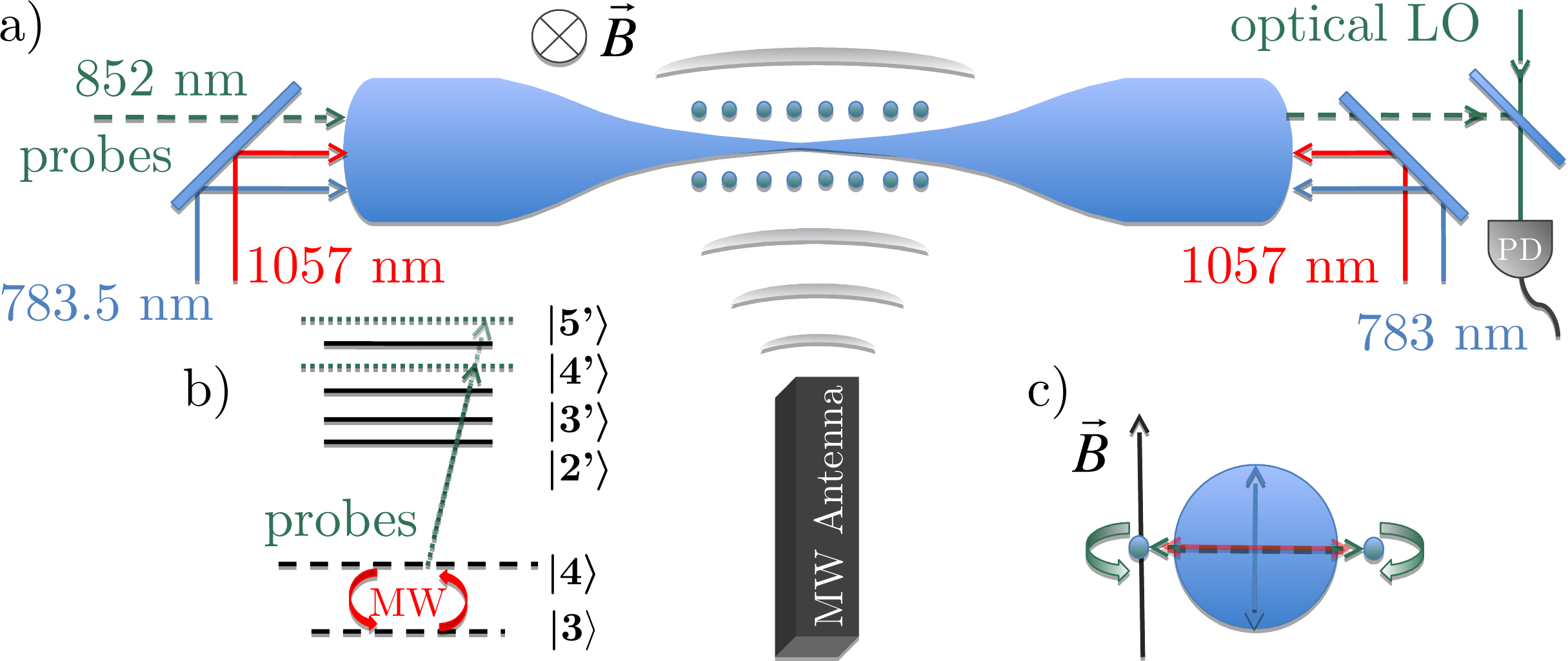}
  \includegraphics[width=\columnwidth]{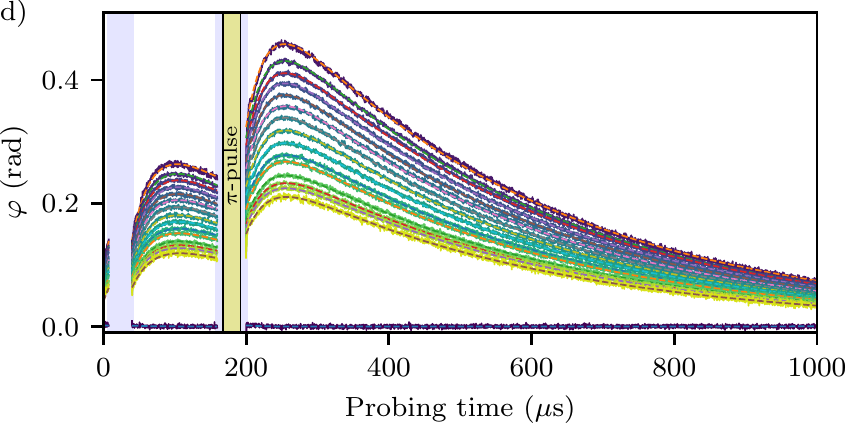}
  \caption{\label{fig:setup}a) Sketch of nanofiber trap setup with
    light fields for trapping and probing and microwave source for
    coherent transfer between clock levels.  b) Relevant atomic energy
    levels for cesium. c) Transverse section of nanofiber with atomic
    trap sites, local probe light polarization, and magnetic bias
    field direction indicated.  d) Dispersive detection of the upper
    clock level population for increasing (bottom to top) atom
    number. Solid lines are averages over typically 300
    realizations. Dashed lines are fits to the model presented in
    equation~(\ref{eq:2}). See main text for state preparation and
    measurement procedure; shaded areas indicate interruption of the
    probing.}
\end{figure}

The experiment begins with loading the nanofiber trapping sites
(\fref[a]{fig:setup}) with cold atoms \cite{JB} from a magneto-optical
trap (MOT) and preparing them in the lower clock level $\ket{3,0}
\equiv (6^2S_{1/2}, F=3, m_F = 0)$ (\fref[c]{fig:setup}) as follows.

We first accumulate the majority of the atoms in the upper clock level
$\ket{4,0} \equiv (6^2S_{1/2}, F=4, m_F = 0)$ using dark state optical
pumping on the $\ket{4} \equiv (6^2S_{1/2}, F=4) \to \ket{4'} \equiv
(6^2P_{3/2}, F=4)$ transition with $\pi$-polarized light propagating
through the nanofiber and external repumping light on $\ket{3} \equiv
(6^2S_{1/2}, F=3) \to \ket{4'}$. A magnetic bias field of
$\SI{3}{\gauss}$ applied along the polarization direction of the blue
trap light defines the quantization axis (\fref[c]{fig:setup}). Using
a resonant microwave Rabi $\pi$-pulse, we transfer the accumulated
atoms into $\ket{3,0}$ before pushing out of the trap any atoms
remaining in level $\ket{4}$ using external blue detuned and
circularly polarized light on the $\ket{4} \to (6^2P_{3/2}, F=5)
\equiv \ket{5'}$ transition.  Finally, we are left with a pure
spin-polarized ensemble of up to $\Natom \sim 1500$ atoms in
$\ket{3,0}$, corresponding to an overall pumping efficiency of about
$65\%$, similar to values reported in\cite{ArnoPRL2013}.

A prerequisite to the characterization of collective atomic states is
a measurement method of the atomic basis state populations. To this
end, building on the versatile and minimally invasive dual-color
dispersive probing method developed in \cite{JB}, we measure the
atomic populations in $\ket{3,0}$ and $\ket{4,0}$:

Two probe light frequencies, symmetrically red- and blue-detuned by
$\SI{62.5}{\mega \hertz}$ with respect to the $\ket{4} \to \ket{5'}$
transition, are sent through the nanofiber~(\fref[b]{fig:setup}). The
differential phase shift $\varphi(t)$ of these fields due to
off-resonant interaction with atoms in $F=4$ is detected via
heterodyne detection with an external optical local oscillator tuned
exactly in the middle between the probe frequencies. This allows for a
shot noise limited dispersive detection of $F=4$ atoms with an overall
homodyne detection efficiency $q = 0.4$.

The probe field polarization is aligned parallel to the red-detuned
trapping light.  Due to the in-quadrature longitudinal field component
of the evanescent probe mode at the location of the atomic trap
potential minima the local polarization is almost circular
($92\%~\sigma^\pm$ and $8\%~\sigma^\mp$) with opposite helicity for
the two 1D-ensembles of atoms trapped on either side along the
nanofiber\cite{mitsch2014,sayrin2015}. Consequently, during probing,
atoms initially prepared in the state $\ket{4,0}$ will be pumped
progressively into the extreme Zeeman levels $\ket{4,\pm 4}$. Here,
the probe interaction occurs predominantly via the strong and closed
$\ket{4,\pm4} \leftrightarrow \ket{5,\pm 5} $ transition, which
increases the interaction strength and reduces decays to the $F=3$
hyperfine level.

The dynamics of the evolution of the mean differential phase shift
$\langle \varphi(t)\rangle$ caused by atoms initially in $\ket{4,0}$
is described by a phenomenological model
\begin{align}
  \begin{split}
    \langle \varphi (t) \rangle & = \varphi_0\cdot\hat{m}(t) \quad \text{with} \\
    \hat{m}(t) & = \left( \beta - (\beta-1)
      \mathrm{e}^{-t/\tau_{\text{at}}}\right)
    \mathrm{e}^{-t/\tau_{\text{loss}}},
  \end{split} \label{eq:1}
\end{align}
where the time constant $\tau_{\text{at}}$ describes the speed of
pumping towards the extreme Zeeman states, $\varphi_0$ denotes the
initial phase shift at $t=0$, and $\beta$ models the increase in
interaction strength during Zeeman pumping. The time constant
$\tau_{\text{loss}}$ is associated to random loss of atomic population
out of $\ket{4}$, caused predominantly by probe-induced heating and to
a much smaller extent to pumping into the $F=3$ hyperfine manifold.

By fitting eq.~\eqref{eq:1} to traces recorded with a varying number
of atoms $N_4$ initially in $\ket{4,0}$, we verify experimentally that
$\tau_{\text{at}}$, $\tau_{\text{loss}}$ and $\beta$ are independent
of the total number of atoms and that they can be calibrated for fixed
experimental conditions (trap and probe powers) such that $\varphi_0$
is proportional to the number of atoms: $\varphi_0 = N_4
\varphi_{\text{eff},1}$.

To determine the population $N_4$ in the $\ket{4,0}$ and the
population $N_3$ in $\ket{3,0}$ in a single realization of a quantum
state, we use the following measurement sequence: First, we probe the
differential phase shift $\varphi(t)$ caused by the atoms in $F=4$ for
a time $\sim 6\tau_{\text{at}}$ (with an \SI{32}{\micro \second}
interruption, see \sref{sec:qnd-meas-weak}); after this time, all
atoms are pumped out of $\ket{4,0}$ and accumulated in the $\ket{4,\pm
  4}$ states. Then, we briefly turn off the probe light for
$\SI{40}{\micro \second}$, apply a microwave $\pi$-pulse to transfer
the population of $\ket{3,0}$ into $\ket{4,0}$ and again probe the
ensemble for $\sim 20\tau_{\text{at}}$ (see \fref[d]{fig:setup}).

We concatenate these successive measurements into a single trace and
the temporal dynamics of the recorded traces are well described by
\begin{align}
  \varphi (t) =
  \begin{cases}
    \varphi_{4}\cdot\hat{m} (t), & 0 \leq t < t_{\text{flip}}, \\
    \varphi_{4}\cdot\hat{m} (t) +
    \varphi_{3}\cdot\hat{m}(t-t_{\text{flip}}), & t_{\text{flip}} \leq
    t.
    \label{eq:2}
  \end{cases}
\end{align}
A linear least squares fit to the measured data yields $\varphi_4$ and
$\varphi_3$, where $\varphi_{4}=\varphi_{\text{eff},1}N_{4}$ and
$\varphi_{3}=\varphi_{\text{eff},1}N_3$. $\varphi_{\text{eff},1}$
denotes the effective differential phase shift caused by a single atom
in the state $\ket{4,0}$, and $N_3$ and $N_4$ are the initial atomic
populations in the $\ket{3,0}$ and $\ket{4,0}$ clock
levels. $t_{\text{flip}}$ indicates the beginning of the $\ket{3,0}$
population measurement.

\section{Projection Noise}

In the following we use this technique to demonstrate, for the first
time, the measurement of quantum fluctuations of ensembles coupled to
a single-mode waveguide.

Using a resonant $\pi/2$ microwave pulse we prepare the ensemble of
atoms in the superposition of the clock states, \mbox{$\ket{\Psi} =
  \bigotimes^{\Natom} \left(\ket{3,0}+\ket{4,0}\right)/ \sqrt{2}$}.
With microwave Ramsey spectroscopy we observe a bare coherence time
$\mathcal{T}_2$ of a few hundreds of microseconds, that can be
increased with spin echo to a few milliseconds, in agreement with the
observations of Reitz et al.~\cite{ArnoPRL2013}.

In order to observe the fundamental quantum fluctuations of the atomic
populations of a coherent superposition state, it is necessary to
reject the shot-to-shot fluctuations in the number of atoms inherent
to state preparation (e.g loading noise), i.e. we need to know the
number of atoms initially in $\ket{4,0}$ and $\ket{3,0}$.

\begin{figure}
  \includegraphics[keepaspectratio, width=\columnwidth]{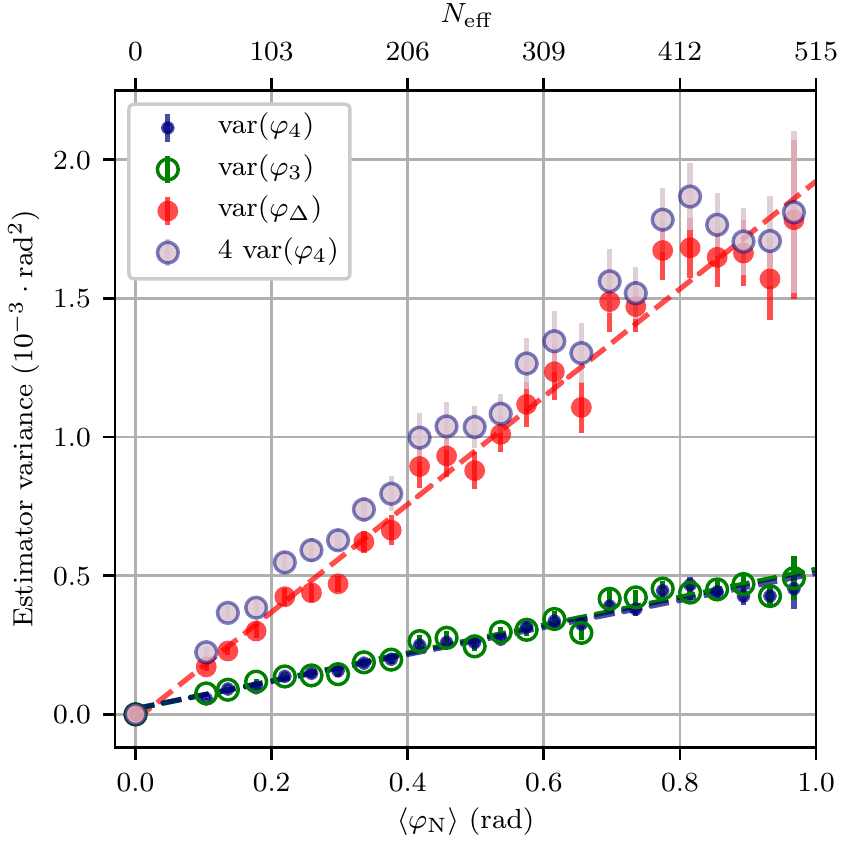}
  \caption{\label{fig:strongscaling} Noise scaling of the atomic
    population estimators.Green and blue symbols are variances of
    $\varphi_{4}$ and $\varphi_{3}$, red filled circles show the
    variance of $\varphi_\Delta$, rose circles show variance of
    $\varphi_{4}$ scaled by a factor of 4.  Error bars are
    statistical, dashed lines are linear fits to the data of same
    color. The top x-axis shows the inferred effective atom number
    $N_{\text{eff}}$.}
\end{figure}

Knowing $\varphi_{4}$ and $\varphi_{3}$, for each single-shot
measurement trace according to eq.~\eqref{eq:2}, we define the phase
estimator proportional to the total atom number $\Natom=N_4+N_3$ as
$\varphi_N = \varphi_{4} + \varphi_{3}$ and for the population
difference $\Delta N=N_4-N_3$ as $\varphi_\Delta = \varphi_{4} -
\varphi_{3}$. For a coherent spin state (CSS) the latter should be
subject to the projection noise: For a CSS the variances of the
populations are $\var{N_{4}} = \var{N_{3}} = \mean{\Natom}/4$ and
$\var{\Delta \Natom} = \mean{\Natom}$. From this we see that for the
projection noise limited state $\var{\varphi_\Delta} -
\delta{\varphi_\text{sn}}^2 = \varphi_{\text{eff},1}^2 \Natom =
\varphi_{\text{eff},1} \mean{\varphi_n}$ scales linearly with
$\mean{\varphi_N}$. Here, $\delta{\varphi_\text{sn}}^2$ denotes
measurement phase noise in the absence of atoms such as shot noise and
detector electronic noise.  In~\fref{fig:strongscaling}, we present
$\var{\varphi_\Delta}$ as a function of $\mean{\varphi_N}$. Each point
corresponds to a bin of around 200 experiments with a similar atom
number. The scaling is linear with a negligible quadratic part over a
wide range of atom numbers. For the experimental realizations with the
maximum atom number, $\var{\varphi_\Delta}$ reaches up to
\SI{40}{\deci\bel} above its value in the absence of atoms
$\delta{\varphi_\text{sn}}^2$, i.e. above the detection noise
alone. From the slope of a linear fit one can extract an effective
phase shift per atom $\varphi_{\text{eff},1} =
\SI{2.0}{\milli\radian}$ and an effective atom number such that $d
\var{\varphi_\Delta} / d \mean{\varphi_N}= \varphi_{\text{eff},1} $
with $\varphi_N = \varphi_{\text{eff},1} N_{\text{eff}}$
\cite{vlatan2015}.  In addition to the negligible quadratic component
which characterizes atom-number dependent technical noise
\cite{monica2010}, the observed ratio of the slopes
in~\fref{fig:strongscaling}, $\var{\varphi_{4}} / \var{\varphi_\Delta}
\simeq 4$ presents a strong confirmation of the quantum projection
noise nature of the presented data, as it is consistent with almost
purely anti-correlated fluctuations of the two clock levels. We find
negligible additional noise when replacing $\pi/2$ pulses by $3 \pi/2$
pulses, which rules out microwave power fluctuations as its origin.

\subsection{$\Natom$ vs $N_\text{eff}$ discrepancy}

We can compare $N_{\text{eff}}$ extracted from the projection noise
scaling to the atom number $\Natom$ obtained via an independent method
based on recording optical pumping transients \cite{JB} which is
robust against important systematic effects such as inhomogeneous
coupling to the probe light.  We observe a systematic difference
$\Upsilon = \Natom / N_{\text{eff}}= 2$, a discrepancy similar to
those reported as well for various other interfaces and treat the
effective atom number as in \cite{vlatan2015}.

We attribute this effect to the coupling inhomogeneity amongst the
trapped thermal atoms located at different positions within the
radially decaying probe field. Such inhomogeneous coupling results in
$\Upsilon = 1 +
{\var{\overline{\varphi_{\text{eff},1}}}}/{\mean{\overline{\varphi_{\text{eff},1}}}^2}
>1 $, where the overscore denotes temporal averaging over the probing
time; variance and mean are taken over the atoms of the ensemble. Due
to the exponential tail of the conservative trapping potential, even
with relatively long averaging times of many tens of radial trap
oscillation periods the mean coupling strength of the individual atoms
still differs. Simulations of the classical trajectories for atoms
with temperatures above $\SI{90}{\micro\kelvin}$ consistently result
in $\Upsilon > 1.5$ even for averaging times beyond $\SI{150}{\micro
  \second}$.

\section[QND measurements]{QND measurements: weak probing and temporal
  dynamics}
\label{sec:qnd-meas-weak}

For a ensemble-waveguide quantum interface of $\Natom$ atoms, ideally
a single optical mode should interface with a single one out of the
$\Natom$ orthogonal atomic modes.

By definition, in nano-optical waveguide structures, electrical fields
vary spatially on the optical wavelength scale, so that inhomogeneous
coupling of the individual quantum systems to the guided light mode is
omnipresent.

Motion of atoms within an inhomogeneous probe field limits the
performance for implementing a quantum interface, since it introduces
temporal dependence on the atomic mode that the optical mode couples
to: information written into such an ensemble cannot be retrieved
perfectly when the atomic modes during the write- and read-process
overlap only partially.

In a spectroscopic setting, the purpose of a weak QND measurement is
to conditionally reduce population noise, while keeping the reduction
of the signal slope $\eta$ due to decoherence low, such that the
signal-to-noise ratio (SNR) improves, i.e.
$\mathrm{var}_\text{cond}\left(\frac{\Delta N}{\eta \Natom}\right) <
\var{\frac{\Delta N}{\Natom}}$, the Wineland criterion for spin
squeezing~\cite{Wineland1992}.  To investigate the influence of atomic
motion and coupling inhomogeneity, in the following we introduce a
framework on how to operate QND measurements under such conditions.

First, we analyze the temporal correlations within single-shot phase
signals $\varphi(t)$: As in the projection noise measurement, atoms
are prepared in $\ket{\Psi}$ and probed for \SI{8}{\micro \second} and
after a \SI{32}{\micro \second} pause again for another
\SI{120}{\micro \second}.

\begin{figure}[tb!]
  {\footnotesize a)}\raisebox{5ex-\height}{
    \includegraphics[keepaspectratio,width=\columnwidth]{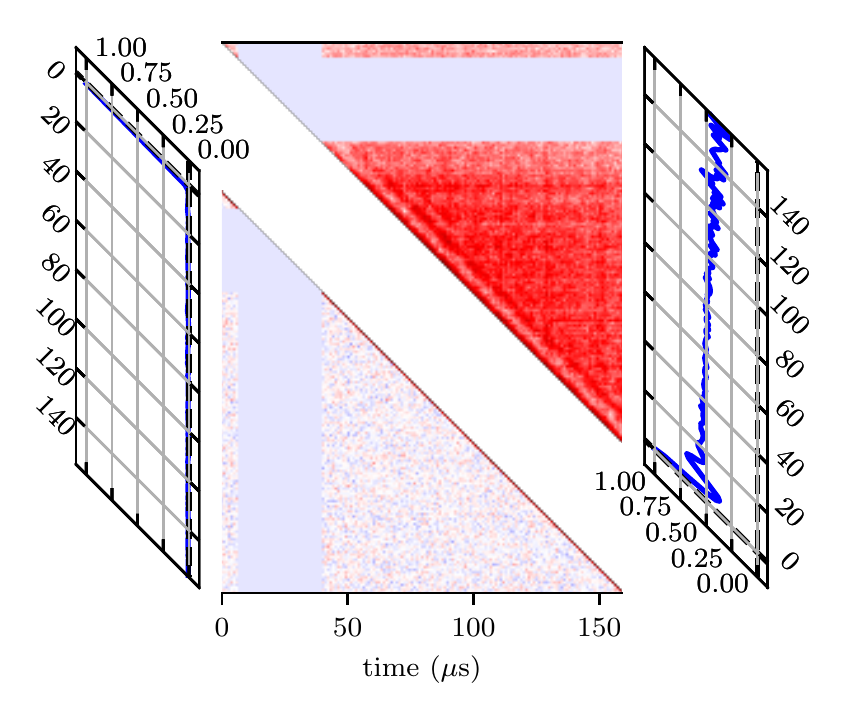}}\\[-5mm]
  {\footnotesize b)}\raisebox{5ex-\height}{
    \includegraphics[keepaspectratio,width=\columnwidth]{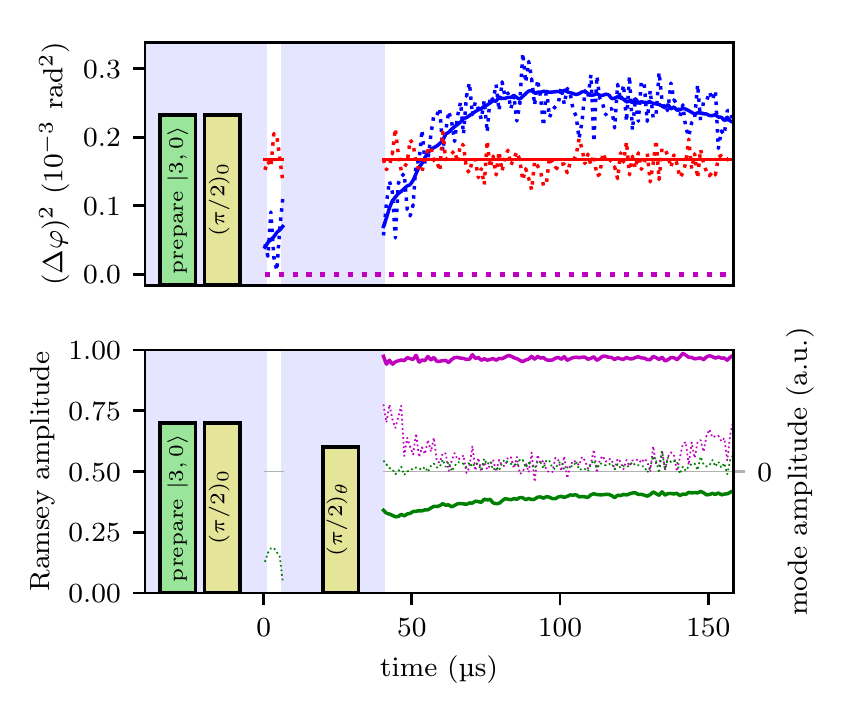}}
  \caption{\label{fig:cov} a)~Constant part $ C_0$ (lower right) and
    linear part $\varphi_\text{N} C_1$ (upper right) of the covariance
    matrix $C=C_0 + \varphi_\text{N} C_1$ for $\varphi_N = \SI{0.64
      \pm 0.03}{\radian} $ (see text); red (blue) denotes positive
    (negative) correlations. Side panels: Correlation coefficient
    obtained by averaging over the minor diagonals of the
    corresponding matrix of correlation coefficients.  b)~Upper panel:
    QND pulse sequence; temporal dependence of the main diagonal
    entries of $C_0$ (constant part, red dots) and of
    $\varphi_\text{N} C_1$ (atom number dependent part, blue
    dots). Solid blue line: $(\varphi_\text{N} \hat m(t)/2)^2 /
    N_\text{eff}$, solid red line: phase shot noise $\delta
    {\varphi_\text{sn}}^2$. Lower panel:~Ramsey-type pulse sequence
    and temporal evolution of the Ramsey fringe contrast in the
    absence (solid purple) or presence (solid green) of a
    \SI{8}{\micro \second}-probe pulse. Dotted traces denote the
    corresponding optimal detection mode function.}
\end{figure}

From each single trace we determine $\varphi_\text{N} := \varphi_3 +
\varphi_4$ according to eq.~\eqref{eq:2}. To eliminate the influence
of atom number fluctuations when loading the MOT, we then subtract
from each sample of the time trace $\varphi(t<t_\text{flip})$ its
expectation value $\frac{\varphi_\text{N}}{2} \hat m(t)$ to obtain the
fluctuations $\delta\varphi(t<t_\text{flip}) :=
\varphi(t<t_\text{flip}) - \frac{\varphi_\text{N}}{2} \hat m(t)$ and
their covariance matrix $C$.

As expected for quantum fluctuations, $C = C_0 + \varphi_\text{N} C_1$
can be decomposed into a constant part $C_0$ and a linear, atom number
dependent part $C_1$.  As in the projection noise measurement, we find
no quadratic contribution scaling with ${\varphi_\text{N}}^2$.  $C_0$
describes the measurement noise in the absence of atoms originating
from the phase shot noise of the low-intensity probe light; as it
represents uncorrelated, constant phase shot noise it can be
approximated as $C_0 = \delta {\varphi_\text{sn}}^2 \cdot \mathbf{1}$
(see lower left part of~\fref[a]{fig:cov}). $C_1$, depicted in the
upper right of~\fref[a]{fig:cov}, describes atomic quantum
fluctuations, such as projection- and partition-noise. The diagonal
elements of $C_1$ follow the square of the evolving mean phase shift
as $(\frac{\varphi_\text{N}}{2} \hat m(t))^2 / N_\text{eff}$. (see
\fref[b]{fig:cov}). After normalizing rows and columns of $C_1$ by
this trace, we obtain the matrix of correlation coefficients (not
displayed) which is of Töplitz-form with minor diagonals as given in
the right inset of~\fref[a]{fig:cov}. In this trace, we observe a
rapid oscillatory decay of the correlations to about half their
initial value with an oscillation period of $\SI{11}{\micro \second}$
and a similar damping time constant.  This can be understood to
originate from the radial oscillation of the atoms within the strongly
anharmonic trapping potential: Due to the inhomogeneous probe field,
within only half a trap oscillation the probed atomic mode has changed
significantly with only minor revivals.

By inserting a $\pi/2$-microwave pulse during the \SI{32}{\micro
  \second} probing break, under otherwise unchanged probing
conditions, we evaluate the loss of Ramsey fringe contrast $\eta(t)$
to determine the ``destructiveness'' of the first \SI{8}{\micro
  \second} probe interval (see \fref[c]{fig:cov}). This central
characteristic is important for protocols such as QND-measurement
based spin-squeezing and quantum metrology applications as $\eta(t)$
is directly proportional to the signal amplitude. The effect of the
atomic motion is also visible in this measurement: a slight increase
of the fringe contrast is observed with increasing probing time, as
atoms affected by the initial decohering light pulse are progressively
leaving the probe region and are replaced by more weakly probed ones.
In this measurement, due to the duration of the microwave-pulse and
the imposed \SI{32}{\micro \second} delay until probing commences, the
damped oscillatory feature seen in the correlation coefficients was
not directly accessible and is surmised to have decayed by the time
when the Ramsey contrast is probed.

For a temporal detection mode $\hat q(t)$, the metrologically relevant
phase resolution is given by the square root of the signal-to-noise
ratio $\mathtt{SNR} \equiv \frac{\hat q^\intercal \cdot s \cdot
  s^\intercal \cdot \hat q}{\hat q^\intercal \cdot C \cdot \hat q}$
with $\hat s(t) = \varphi_N \, \hat m(t)\,\eta(t)$ being the
decoherence-affected signal strength and $C$ the covariance matrix.
Knowing both $\eta(t)$ and $C$, we can directly obtain the optimal
temporal probe mode function that maximizes the SNR for both the
spin-squeezing pre-measurement and the spectroscopic measurement using
a matched filter \cite{Turin1960} as $\hat q_\text{opt}(t) = C^{-1}
\cdot \hat s(t)$ (see \fref[c]{fig:cov}). However, in this experiment,
due to the rapid decay of the correlations and the low optical density
of the present ensemble, even with the optimal detection mode, using a
QND pre-measurement we cannot improve spectroscopic resolution
compared to a conventional Ramsey sequence.

Several technical measures can push the system deeper into the quantum
regime. Utilizing the hyperfine coherence between stretched Zeeman
levels, the optical depth for the pre-measurement can be increased by
a factor of three.  Working with faster coherent operations and
stronger probing the measurement can outrun the thermal motion induced
decay of correlations. Last, but not least, the decay of correlations
can be greatly reduced by cooling the atoms into the motional ground
state of the lattice sites~\cite{Albrecht2016,Ostfeldt2017,Meng2017}.

Given the inherent inhomogeneous coupling between light and atoms in
nanoscale traps it is relevant to discuss the limits posed by quantum
mechanics on the performance of such inhomogeneous dispersive
atom-light interfaces.  As in the case of homogeneous QND-coupling the
noisy quantum backaction onto the atomic coherence is mediated by
residual spontaneous scattering of probe light. In the inhomogeneous
case the change of motional state due to photon recoil leads to
additional noise, since the coupling generally depends on the trap
level. The probability for leaving the motional ground state in a
single scattering event is suppressed by the Lamb-Dicke parameter
$\eta^2 = \omega_{rec}/\omega_{trap}$, implying that strong enough
confinement and low temperature allow to keep the extra noise far
below the inevitable dephasing. We note that engineering the
dissipation by channeling a sizable fraction of spontaneous emission
into the fiber guided modes mediates long-range spin-dependent
interaction between the atoms. This changes the simple picture and
opens a rich field of many-body physics for study \cite{njpChang,
  Alejo_PRL, Ana_PRX}.

\section{Conclusion} 
We have presented an experimental analysis of atomic quantum noise
detection for ensembles trapped in the evanescent field of an optical
nanofiber and identified the main challenges to be overcome for truly
quantum enhanced applications.  Our work paves the way to entanglement
generation of a few atoms using solely the light propagating in a
nano-optical waveguide interface. As a single nanofiber system is the
host of two mesoscopic one-dimensional atomic ensembles, it offers the
exciting prospect of a joint QND (Bell) measurement of the two
ensembles, a well identified goal for EPR entanglement and
teleportation protocols based on two-mode squeezing. For these
applications an integrated fiber cavity can significantly enhance the
effective optical depth of the ensemble
\cite{AokiNanofiber,Ruddell2017} and improve on the degree of
achievable squeezing.

\begin{acknowledgments}
  This work has been supported by the ERC grant INTERFACE (grant
  no.~ERC-2011-ADG~20110209), the US ARO grant no.~W911NF-11-0235 and
  the EU grant SIQS (grant no.~600645).
\end{acknowledgments}

\bibliographystyle{apsrev4-1} %needed to suppress bibtex-'Warning--jnrlst'
% \label{LastBibItem}
\bibliography{articlebib}

\end{document}